%% file: ms.tex
\documentclass[12pt, preprint]{aastex}
 \newcommand{\Msun}{\mbox{M$_{\odot}$}}
 \newcommand{\Rsun}{\mbox{R$_{\odot}$}}
 
 \newcommand{\ltsimeq}{\raisebox{-0.6ex}{$\,\stackrel
        {\raisebox{-.2ex}{$\textstyle <$}}{\sim}\,$}}
 
\shorttitle{The Intermediate Polar EI UMa}
\shortauthors{Reimer et al.}

\begin{document}

\title{The Intermediate Polar EI UMa: A Pre-Polar Cataclysmic Variable}

\author{Tamara W. Reimer}
\affil{Department of Astronomy, San Diego State University,
    5500 Campanile Drive, San Diego, CA 92182-1221}
\email{tamara.reimer@gmail.com}

\author{William F. Welsh}
\affil{Department of Astronomy, San Diego State University,
    5500 Campanile Drive, San Diego, CA 92182-1221}
\email{wfw@sciences.sdsu.edu}

\author{Koji Mukai}
\affil{Astrophysics Science Division (Also University of Maryland, 
Baltimore County), NASA Goddard Space Flight Center, Code 662, Greenbelt, 
MD 20771}
\email{mukai@milkyway.gsfc.nasa.gov}

\author{F. A. Ringwald}
\affil{Department of Physics, California State University, Fresno, 2345 
East San Ramon Avenue, M/S MH37, Fresno, CA 93740-8031}
\email{ringwald@csufresno.edu}

% -----------------------------------------------------------------
\begin{abstract}

We present optical and X-ray time-series photometry of EI~UMa that reveal
modulation at 746 and 770~s, which we interpret as the white dwarf spin 
and spin-orbit sidebands. These detections, combined with 
previous X-ray studies, establish EI~UMa as an intermediate 
polar. We estimate the mass accretion rate to be $\sim 3.6 \times 10^{17}
$~g s$^{-1}$, which is close to, and likely greater than, the critical rate 
above which dwarf nova instabilities are suppressed. We also estimate the 
white dwarf to have a large magnetic moment $ \mu > (3.4 \pm 0.2) 
\times 10^{33}$~G~cm$^3$. The high mass accretion rate and magnetic moment 
imply the existence of an accretion ring rather than a disk, and along 
with the relatively long orbital period, these suggest that EI~UMa 
is a rare example of a pre-polar cataclysmic variable.

\end{abstract}


% -----------------------------------------------------------------
\section{Introduction}

Cataclysmic variables are binary star systems in which gas is 
gravitationally stripped from a companion star and accreted onto a white 
dwarf, typically via an accretion disk. In intermediate polar systems
(also known as ``DQ Her stars'') the accretion disk is disrupted by the 
magnetic field of the white dwarf.  Transferred material forms a disk at 
large radii, but follows the magnetic field of the white dwarf when inside 
the white dwarf's magnetosphere. The rotation of the white dwarf and the 
magnetically channelled accretion flow creates periodic modulations in 
the observed flux. \citet{patterson94} lists six observational 
criteria used to identify an intermediate polar: 1) a stable optical 
period $P_{spin}$ that is less than the orbital period; 2) a stable X-ray 
period at or near the optical period; 3) pulsations in the He II emission 
lines; 4) circular polarization; 5) a sideband period in the optical 
and/or X-ray, usually on the long period side of the main signal; and 6) 
a very hard X-ray spectrum with low-energy absorption.  A cataclysmic 
variable need not exhibit all six of these characteristics to be an 
intermediate polar (henceforth abbreviated as IP).

EI Ursa Majoris, also known as PG 0834 + 488, was discovered in the 
Palomar-Green Survey of objects with ultraviolet excesses, and 
subsequently found to be a cataclysmic variable \citep{green82}.  
H$\alpha$ radial velocities show the system to have an orbital period of 
6.434~h \citep{thorstensen86}.  Early observations suggested this  
object exhibits some of the characteristics of an IP. Green et al. 
observed that the system has an unusually strong He II 4686 
\AA\ line. \citet{cook85} noted that the system exhibits a strong hard 
X-ray flux. The early X-ray observations failed to show any stable 
periodic modulation however, leading Cook to suggest that EI UMa was 
instead a dwarf nova. This classification currently persists (e.g. 
\citealt{ritterkolb03}), despite the lack of observed outbursts for this 
system (see Figure 1). Recent X-ray observations support classification of 
EI~UMa as an IP, and not as a dwarf nova. \citet{pandel05} argue for the 
IP status based on XMM-Newton X-ray spectra. They also found EI~UMa to 
have a low UV luminosity, and suggest this is a result of the missing 
inner accretion disk. \citet{baskill05} claim EI~UMa is an IP based on 
its ASCA X-ray spectral characteristics, and most importantly, the 
discovery of a 741.6 $\pm$ 5.4~s periodicity with an amplitude of 8 $\pm$ 
1\%.

EI~UMa has thus satisfied Patterson's conditions 2 and 6 for being an IP. 
If a stable optical period were found, the case for EI~UMa's IP status
would be much strengthened. We therefore collected 84 hours of optical 
photometry (see \S2) to search for oscillations near the X-ray period 
discovered by Baskill et~al. We present the detection of such optical 
oscillations in \S3 of this paper, and present a second detection of the 
X-ray period using archival XMM-Newton data in \S4. In \S5 we discuss our 
search for longer timescale phenomena, including ellipsoidal modulations 
and dwarf nova outbursts. We examine EI~UMa as an IP in \S6, estimating 
its distance, mass accretion rate, and magnetic field strength, and we 
summarize our results in \S7.

% -----------------------------------------------------------------

\section{Optical Observations}

We observed EI UMa on 16 nights between 2003 February and 2005 December 
using the 1-m telescope and CCD at San Diego State University's Mount 
Laguna Observatory. An observation log is presented in Table~1.  
Observations on the first nine nights looked for modulations on the spin 
period and harmonics; this set included five nights of V-band data, two 
nights of U-band data, and two nights of B-band data. Observations during 
the last seven nights looked for variability on the longer orbital 
timescale and used an R-band filter. Images were bias-subtracted and 
flat-field corrected using standard IRAF data reduction routines, and 
light curves were extracted by performing aperture photometry with the 
APPHOT package. A large aperture of 4.8 arcseconds was chosen for the 
R-band observations to minimize the night--to--night changes in 
comparison star light curves. The calibration of the R-band light curves 
is accurate to $\sim$2\%, based on the relative flux of the check star. 
\citet{misselt96} gives BVR magnitudes for EI~UMa (B=14.90, V=14.78, and 
R=14.55) and nearby field stars. Differential light curves from the first 
four nights use Misselt's star~1 (M1; also star 21 in 
\citealt{henden97}). The remaining 12 light curves use Misselt's star~3 
(M3; star 10 in \citeauthor{henden97}), and Misselt's stars~1 and 4 were 
used as check stars. Nightly mean V-, B-, and R-band magnitudes for 
EI~UMa are presented in Table 2. Comparison magnitudes in the U-band were 
not available. Figure 2 shows our U, B, and V differential flux light 
curves. The most obvious feature is the incessant flickering, a defining 
characteristic of accretion--powered objects.  The RMS of the 
red noise
flickering in the differential fluxes is typically $\sim$5\%. This RMS has 
not been corrected for observational noise, and thus is slightly 
overestimated.

% -----------------------------------------------------------------
\section{The Optical Periodicities}

\subsection{Time Series Analysis}

Power spectra were computed to search for periodicities near the 
previously detected X-ray periodicity ($741.6 \pm 5.4$~s; 
\citealt{baskill05}). Before Fourier transformation, the light curves were 
detrended to minimize low frequency red noise leakage. We used a sliding 
boxcar to remove power on timescales longer than 45 minutes. Other 
detrending methods were tried as a check for robust periodicity; this 
included subtracting off only the mean or a linear fit to the light curve. 
Power spectra for 2003 February 1, 2, and 21 each show a significant 
signal near 745~s. The Lomb--Scargle periodogram \citep{press03}
yields a false alarm probability of $4 \times 10^{-18}$ for the Feb~21 
signal.  Although this formal probability is meaningless in the presence 
of red noise, much of the red noise is removed by the prewhitening 
(detrending) of the light curve, and the extremely small false alarm 
probability shows that the detection is highly significant. Six additional 
nights exhibit weaker, but measurable, signals; these signals are not all 
at the same period, but range between 747 and 815~s. The average 
amplitude of the periodicity is $\sim$ 1\%. The remaining seven light 
curves show no detectable periodicities in the period range between 700 
and 850~s. No significant shorter period signal was detected in any light 
curve, indicating no strong harmonics are present.  For nights when a 
periodicity was not detected, amplitude limits were obtained by injecting 
``tracer'' sinusoids of different frequencies and phases into the light 
curves.  The detection limit threshold was defined as the amplitude at 
which two out of five injected sinusoids were no longer apparent above 
the noise in the power spectrum. Figure~3 shows the stronger detections, 
Figure~4 shows examples of non-detections and tentative detections, 
and the periods and amplitudes are given in Table~2.

\subsection{Spin and Sideband Periods}

The observed optical periods fall into three groups: 747~s,  
760--770~s, and 813~s. The 747~s periodicity is seen only in the two 
U-band observations (separated two years apart). This agrees 
with the $741.6 \pm 5.4$ s X-ray periodicity seen by \citet{baskill05} and 
we associate these with the spin period of the white dwarf. Combining the 
U-band and X-ray measurements, we derive a weighted average spin period 
of $745.7 \pm 2.7$~s.  We note that the two U-band signal detections are 
not convincing by themselves, but are statistically significant when 
viewed in combination with the X-ray and the other optical signals.

We identify the stronger 760--770~s oscillation as the first-order 
sideband period $P_{side}$ given by $P_{side}^{-1} = P_{spin}^{-1} - 
P_{orb}^{-1}$. Using the orbital period of $6.434 \pm 0.007$~h 
\citep{thorstensen86}, the sideband period for this system should be 
$770 \pm 3$~s.  Our light curves have just enough resolution to 
distinguish the spin and sideband periodicity, as sinusoid injection 
simulations give a resolution of $\> \sim $20~s. The discovery of the 
optical spin and sideband periodicities (criteria (1) and (5) in 
\citealt{patterson94}) combined with the X-ray periodicities found by 
\citet{baskill05} and also in this work (see \S4) provide compelling 
evidence that EI~UMa is an intermediate polar.

Intermediate polars are known to exhibit other sideband periods, so we 
searched for --- but failed to confirm --- periodicities at other 
sidebands. The third--order sideband $P_{side}^{-1} = P_{spin}^{-1} - 3 
P_{orb}^{-1}$ at $821 \pm 3$~s is in the vicinity of the 813~s period seen 
on two, and possibly three, occasions. However, the agreement is not good 
enough to firmly associate these with the third sideband. The bottom three
panels in Fig.~4 show the possible third--order sidebands detections.

The presence of an orbital sideband requires a re-processing site that is 
locked in the rotating binary frame, such as the bright-spot 
\citep{hassall81} or the secondary star \citep{pattersonprice81}. 
Interestingly, the spin period is seen only in our U-band light curves 
and the sidebands are seen only in the V- and R- bands, where reprocessed 
radiation is more likely to be seen.  Systems that similarly 
exhibit strong sideband periods with spin periods that are hard to detect 
at optical wavelengths include V1223~Sgr \citep{osborne85} and YY Dra 
\citep{patterson92}.  In the latter system, the U-band is the only 
optical wavelength in which the quiescent spin period is even marginally 
visible \citep{haswell97}.

\subsection{Signal Amplitude and Stability}

Since the lightcurves of cataclysmic variables are subject to flickering
and apparent short-lived periodicities, it is important to establish the
stability of any potential spin/ sideband period signal 
\citep[e.g., see][]{warner89}. Periodicities were detected in nine out of 
sixteen light curves over a timespan of almost three years, and there is 
no evidence that the periods change.  This suggests that the 
periodicities are stable, but that the signal amplitude is variable and 
sometimes drops below our detection threshold.  The relative amplitudes 
of detected periodicities also vary, ranging from 0.4 to 1.3\%.  The 
highest relative V-band amplitude was 1.3\% while the lowest V-band 
non-detection limit was 0.6\%; therefore, there is at least a factor of 
two change in fractional pulse amplitude. Though we have only a few 
detections, there is a tendency for the relative pulse amplitude to 
decrease with increasing system brightness. Enhanced accretion then does 
not result in a corresponding increase in the pulse amplitude. This 
suggests that the pulsations arise in a region physically distinct from 
the accretion disk --- a scenario consistent with the intermediate polar 
interpretation.

% -----------------------------------------------------------------

\section{Periodicity in Archival XMM-Newton Data}

In their spectral analysis of EI~UMa using XMM-Newton data, 
\citet{pandel05} comment that they did not find any clearly periodic 
signal in the X-ray or UV data.  To further investigate the presence of 
an X-ray periodicity, we re-examine the archival XMM-Newton data. To 
enhance detection, we have separately analyzed the soft (0.3-2 keV) and 
hard (2-12 keV) bandpasses.  

The observations were obtained on 2002 May 10 and 2002 Nov 2.  We 
concentrate on the medium spectral resolution data obtained with the 
European Photon Imaging Camera (EPIC), which consists of one PN and two 
MOS CCD arrays. We have reprocessed the observations using SAS 7.0.0 and 
extracted light curves in 16~s bins. The May observation was affected by 
soft proton flares, which significantly increase the particle background 
rate; these time intervals were eliminated. The background was 
well-behaved for the November observation.

Because the observation duration is shorter for the PN than for the MOS
cameras, we primarily used the combined MOS-1 + MOS-2 light curves
for period search (though the PN data yield similar results).  The 
results are presented in Figure~5.  A periodicity consistent with 745~s 
is clearly detected in the low energy light curve of the November 
observation, and marginally detected above 2 keV. Since the spin 
modulations in IPs are energy dependent, with higher amplitudes observed 
at lower energies, this is not surprising. Strong flickering in the May 
observations produced red noise that partially obscured the periodicity, 
but the signal remains present. We cannot claim independent 
detection of the 745~s signal in the hard bandpasses or even the soft 
bandpass in the May data, but the periodograms are consistent with 
the presence of such a signal, and the November soft bandpass clearly 
shows a detection. Note that because the XMM-Newton light curves are of 
shorter duration than the optical light curves, the frequency resolution 
of the X-ray power spectra is considerably worse. The power spectra shown 
in Fig.~5 have been oversampled and the true period resolution is $\sim$ 
50~s. The detections are therefore consistent with both the spin and the 
sideband periods, although we strongly favor the spin interpretation.

% ------------------------------------------------------------------
\section{Long Period Brightness Fluctuations}

\subsection{Lack of Orbital Modulation}

To refine the orbital period and tighten the constraints on the system 
inclination (nominally 23 degrees --- \citealt{thorstensen86}), we 
searched for orbital modulation in the 2005 Nov--Dec set of R-band 
photometry. Given the 6.434~h orbital period of the binary, the secondary 
star should be relatively bright. With EI~UMa designated a dwarf 
nova, the accretion luminosity should also not overwhelm the secondary 
star light. We used IRAF's phase dispersion minimization algorithm to 
search for any modulation.

Using the combined set of light curves, no significant signal was seen 
near 6.434 h, at half or double this period, or anywhere in the range of 1 
to 20 hours.  Removing the mean before combining each night's observation 
produced the same null result, as well as looking at each light curve 
individually. Six light curves are shown in Fig.~6, folded on 
the 6.434~h period; while slow flickering is noticeably present (also 
see Fig.~7), no periodic signal was found. The lack of any 
ellipsoidal variations demands a low inclination, consistent with 
the findings of \citet{thorstensen86} and the IUE spectral modeling of 
\citet{urbansion06}. Furthermore, the lack of any apparent bright-spot 
also suggests a low inclination, assuming the bright-spot is 
azimuthally elongated and radiates preferentially in the orbital 
plane. An alternate explanation for the lack of detection of the 
bright-spot is that its light is overpowered by a luminous disk. A bright 
disk would also explain the lack of any secondary star attributes like 
ellipsoidal variations or spectral lines, a result that is otherwise hard 
to explain given that the secondary is likely a K star 
\citep{thorstensen86}.  A high mass-transfer rate through the disk would 
imply that EI~UMa is more akin to a novalike cataclysmic variable than a 
dwarf nova. We return to this point in \S6.

\subsection{Flare Event and Dwarf Nova Status} % ----------------------

Fig.~7 shows the seven nights of R-band observations taken between 2005 
Nov~18 and Dec~05. The general shape of the light curve is that of a 
broad, 0.65 mag amplitude ``slow hump'' spanning more than 20 days. If we 
define the R-band quiescence level to be 14.55 mag as given by 
\citet{misselt96}, the system flux is slightly elevated on Nov 18 and 
returns to the quiescent level by Dec~05. On Nov 20, on top of this slow 
variation, is a sharp ``fast flare'' lasting several hours. In nearly 90 
hours of photometry we see no other feature that resembles the fast flare 
event. The amplitude of the fast flare cannot be determined because 
the peak was not observed, but a lower limit is 0.27 mag, and the 
average rate of decay was 0.09~mag/h over the observed 3~h decline. The 
brightness difference between the minimum on Dec~05 and the peak of the 
flare on Nov~20 was $>0.80$ mag (a factor of $>2.1$ change in 
flux).

We briefly entertain the hypothesis that the slow hump is a very low 
amplitude dwarf nova outburst. \citet{warner95} gives an empirical 
outburst decay time as a function of orbital period (his equation 
3.5): 
\begin{equation}
\tau_{decay} = 0.53\ P_{orb}^{0.84}(h)\ \mathrm{d\ mag^{-1}}.
\end{equation}
Using the 6.434~h orbital period gives a decay timescale of 2.5~d per mag, 
and for a small 0.65~mag amplitude outburst, the duration is expected to 
be only $\sim 1.6$~d --- clearly inconsistent with the observed $> 10$~d 
decay of the slow hump. If we include the peak brightness of the flare 
observed on Nov~20, a 0.80~mag amplitude drop in brightness would still 
usually take only $\sim 2$~d. 

Is EI~UMa a dwarf nova? AAVSO and VSNet light curves over the last 13 
years show substantial day to day variability of 0.5~mag or more, and a 
slower variation of approximately 1~mag over a timescale of 5 years, 
but do not reveal any obvious dwarf nova outbursts (see Figure 1). We 
searched SIMBAD's list of relevant literature for observations of the 
mean outburst amplitude, duration, recurrence time scale, etc., but 
failed to find such information. The evidence suggests that EI~UMa's 
classification as a dwarf nova is suspect.

Some IP systems have exhibited dwarf nova outbursts (e.g. EX~Hya, 
GK~Per). Of particular note are three systems that have exhibited short 
($<2$~d), low-amplitude ($\lesssim2$~mag) outbursts: V1223~Sgr 
\citep{amerongen89}, TV~Col \citep{hellierbuckley93}, and V1062~Tau 
\citep{lipkin04}. It has been suggested that these short outbursts may 
not be normal disk instability dwarf nova outbursts, but instead are 
enhanced mass transfer bursts from the secondary star (see 
\citealt{hellierbuckley93}; \citealt{hellier97}; \citealt{lipkin04}). 
The outburst timescales for TV Col \citep{hellierbuckley93} are especially 
similar to what we have observed with EI~UMa. TV Col showed a rapid 
$\sim2$ mag flare over several hours in a blue bandpass, followed by a 
much slower, low amplitude decay of $\sim$3 -- 5 d before the system 
returned fully to pre-outburst magnitude levels. The flare exhibited a two 
stage decay: a sharp decline of 0.6 mag in 40 min, followed by slower 
decline of 0.6 mag over 5 h. The 
rate of EI~UMa's fast flare decay ($\sim0.09$~mag/h) is comparable with 
the rate for the latter part of TV~Col's decay ($\sim0.12$~mag/h). By 
analogy with TV~Col, we suggest that the 2005 Nov--Dec event may have 
been a mass transfer burst from the secondary star. Spectroscopic 
observations of the bright-spot and stream emission during a flare could 
test this hypothesis. In Table 3 we list the outburst properties of the IP 
systems known to exhibit outbursts, and add EI UMa to the list for 
comparison.

% -----------------------------------------------------------------
\section{EI UMa as an Intermediate Polar}

Having made the case that EI~UMa is an IP, we now examine the properties 
of EI~UMa and how it compares with other systems in its class. Of the 26 
or so \footnote{For example, see 
http://asd.gsfc.nasa.gov/Koji.Mukai/iphome/iphome.html} IPs, EI~UMa's 
12.4~min spin period is quite ordinary, being close to the middle of the 
distribution of spin periods (0.5~min -- 120~min). EI~UMa's 6.43~h orbital 
period is longer than about 80\% of cataclysmic variables, and is more 
typical of higher accretion rate systems than of dwarf novae 
\citep[e.g.~see][]{shafter92}.  The spin-to-orbital period ratio $P_{spin} 
/ P_{orb}$ is 0.03, smaller than the equilibrium period ratio of $\sim 
0.1$ for typical IPs or 0.07 for EI~UMa itself \citep{king91}.
Thus it is clear that EI~UMa is not in spin equilibrium.

The mass of the secondary star can be estimated from the semi--empirical 
donor star sequence work by \citet{knigge06}. Extrapolating to just 
beyond Knigge's donor star sequence table, an orbital period of 6.43~h 
has $M_{2} = 0.81\pm0.05~\Msun$, $R_{2} = 0.76\pm0.03~\Rsun$, and a 
spectral type of approximately K4; the errors here are reasonable 
estimates. It is required that the mass ratio $q \equiv \frac{M_2}{M_1} 
\ltsimeq 1$ for dynamically stable mass transfer, so this places a lower 
limit on the mass of the white dwarf: $M_{1} \gtrsim 0.81\pm0.05~\Msun$. 
Combined with the white dwarf mass-radius relation \citep{nauenberg72}, 
the upper limit on the white dwarf radius would be $R_1 = 7000$~km. A 
firm upper limit for the mass of the white dwarf is the Chandrasekhar 
mass limit, but a more reasonable upper limit of $M_1 \lesssim 1.2~\Msun$ 
will at times be used for illustrative purposes in the following 
discussion.  

\subsection{Distance} %------------------------------------------

\citet{thorstensen86} states that EI~UMa's spectrum showed no absorption 
features except for ``a very marginal feature at Na~D \ldots
suggesting that less than 40\% of the light is from the companion.''
Using this 40\% limit and a mean magnitude of V=14.5 mag, the secondary
star has an apparent magnitude $V_{2} > 15.5$ mag. The Galactic neutral
hydrogen column density\footnote{from NASA's HEASARC nH Tool at
http://heasarc.gsfc.nasa.gov/cgi-bin/Tools/w3nh/w3nh.pl \citep{dickey90}} 
in the direction of EI~UMa is $n_{H} = 3.35 \times 10^{20}$ cm$^{-2}$,
which gives an extinction of $A_{V} < 0.187 \pm 0.003$ mag using the
ROSAT-derived relation between soft X--ray absorption and optical
extinction, as given by \citet{cox00}. This extinction is highly 
uncertain, but its inclusion or omission does not substantially affect the 
discussion that follows. 

The absolute magnitude of a K4~V star is $\sim$ 7.2 \citep{gray05}, so we 
can estimate a crude minimum distance of $>420$ pc. A much better 
distance estimate can be made by using the 2-MASS measured 
K=13.534 mag \citep{skrutskie06} and the revised K-band Barnes--Evans 
surface brightness relation as given by \citet{beuermann06}. This yields 
an upper limit distance of $< 810$~pc, assuming zero extinction in the 
K-band and that all of the K-band light and 40\% of the V-band light comes 
from the secondary star. Beuermann also gives the K-band surface 
brightness in terms of spectral type, which is independent of the 
observed (V--K) color. Using a K4 spectral type yields a distance of 
745~pc. Agreement between the K-band and spectral type derived distances 
can be made if the contribution of the secondary star is 25\%. A distance 
of $\sim$ 670~pc is derived if the V--band surface brightness is used 
instead, but the V-band is much more heavily contaminated by sources of 
light other than the secondary star, thus making the estimate less 
reliable. Finally, using Knigge's semi-empirical K-band absolute 
magnitude ($M_{K} \approx 4.2$) based on the orbital period, a distance 
of 740~pc is obtained, in good agreement with the above values. Despite 
the good agreement, we will treat these distance estimates with 
appropriate caution and adopt a distance of $750^{+100}_{-200}$ pc in the 
discussion that follows; note that these error bars represent upper and 
lower limits, rather than $1\sigma$ uncertainties.

\subsection{Mass Transfer Rate} %--------------------------------

\citet{baskill05} measured a 0.8--10 keV X-ray flux of $30.0 \times 
10^{-12}$ erg s$^{-1}$ cm$^{-2}$; when combined with the above distance, 
this gives a system X-ray luminosity of $L_{X} \approx 1.0^{+0.3}_{-0.5} 
\times 10^{33}$ erg~s$^{-1}$. Using the conversion from $L_{X}$ to mass 
accretion rate given in \citet{warner96} ($L_{X}=2.8 \times 10^{15} 
\dot{M}$ erg~s$^{-1}$) yields $\dot{M} = 3.6^{+1.0}_{-1.7}\times 10^{17}$ 
g s$^{-1}$. A second method of finding $\dot{M}$ can be made if we assume 
the secondary star contributes $\sim$25\% of the V-band light, and thus 
the absolute magnitude of the disk is $M_{V disk} = 5.4^{+0.7}_{-0.2}$. 
Using the $M_{V}$---$\dot{M}$ relations in Fig.~2 of \citet{smak89}, the 
mass accretion rate is $\gtrsim 4.0 \times 10^{17}$ g s$^{-1}$, in good 
agreement with the mass accretion rate based on the X-ray luminosity. 
This range of mass accretion rates is far greater than the boundary layer 
$\dot{M}$ estimated by \citet{pandel05}, due to the much smaller distance 
of 100~pc assumed in that study.

The disk instability model has been relatively successful at reproducing
observations of dwarf nova outbursts \citep[e.g. see][]{warner95}.  The 
model predicts a critical mass accretion rate, $\dot{M}_{crit}$, above 
which the disk remains in a hot, high viscosity, stable state and the 
cycle of dwarf nova outbursts is suppressed. This limit has been 
observationally confirmed \citep[][specifically Fig.~3.9]{warner95}.  The 
expression for the $\dot{M}_{crit}$ is given by \citet[eq.~37]{warner96}:
\begin{equation}
\dot{M}_{crit} = 8.26 \times 10^{15} \left(\frac{f}{0.3}\right)^{21/8}
\left(\frac{\alpha_{hot}}{0.3}\right)^{3/10}
\left(1+q\right)^{7/8} P^{7/4}_{orb} \ \ \ g \ s^{-1},
\end{equation}
where $\alpha_{hot}$ is the usual viscosity parameterization and
$f \equiv r_{disk}/a$ is the ratio of the outer disk to the orbital
separation. The value of $f$ in intermediate polars is probably in the
range 0.2--0.3 \citep{warner96}, somewhat smaller than in the 
non-magnetic systems; this is because magnetic truncation 
of the inner disk means less angular momentum needs to be transferred 
to the outer disk, and hence the outer disk will not grow as large 
\citep{angelini89}. For EI~UMa, $\dot{M}_{crit} = 1$--$4 
\times 10^{17}$~g~s$^{-1}$.  If we use the $\dot{M}_{crit}(P)$ 
expression from \citet{shafter92}, we get a similar estimate of 
$\dot{M}_{crit} \sim$ 1--2$\times 10^{17}$~g~s$^{-1}$. Thus, the mass 
accretion rate we estimate ($3.6 \times 10^{17}$~g~s$^{-1}$) is close to, 
and most likely larger than, the critical mass accretion rate. Such a 
high $\dot{M}$ is expected to produce a highly luminous disk, explaining 
why we do not see ellipsoidal variations or a bright spot. The high 
$\dot{M}$ also explains why dwarf nova outbursts have not been observed: 
they simply do not occur, and EI~UMa is not a dwarf nova.

\subsection{Magnetic Moment} %----------------------------------

In order for a cataclysmic variable to be an IP, the white dwarf magnetic 
moment $\mu$ must be strong enough to disrupt the accretion disk and 
channel the accretion flow, but not so strong as to completely prevent 
formation of a disk.  \citet{warner96} 
gives this condition as 
\begin{equation}
\frac{\mu_{33}}{\dot{M}_{16}^{\frac{1}{2}}} < 7.04 \times 10^{-2}
\frac{(1+q)^{\frac{7}{12}}}{q^{\frac{3}{4}}} M_1^{5/6} P_{orb}^{7/12},
\end{equation}
where $\mu_{33} = \mu / 10^{33}$ G cm$^3$ and $\dot{M}_{16} = \dot{M} / 
10^{16}$ g s$^{-1}$. In addition, the interaction between the white dwarf 
and secondary star magnetic fields produces a spin-synchronizing torque of 
strength $\zeta$. To be an IP, the magnetic moment must not be so strong 
as to force synchronous rotation, and this leads to a second condition 
given by \citet{warner96}: 
\begin{equation}
\frac{\mu_{33}}{\dot{M}_{16}^{6/5}} < 0.128 (\zeta_{max})^{7/5} (M_1^2)
(1+q)^{7/5} (P_{orb}^{-7/20}),
\end{equation}
where $\zeta_{max} = 1.25.$ Figure 8 shows the disk and synchronization 
conditions for the $P_{orb}$ and $M_1$ range of EI~UMa. 

\citet{warner96} notes that IPs observationally tend to cluster near the 
$\zeta = \zeta_{max}$ line, as shown in Figure 7 of that work. Assuming 
this is also true for EI~UMa, the range of mass accretion rates given by 
the X-ray luminosity yields $\mu > (3.4 \pm 0.2) \times 10^{33}$~G~cm$^3$ 
and $B_1 > 9.9 \pm 0.6$~MG. Typical white dwarf field strengths among IPs 
are $1 - 10$~MG \citep{hellier01}, which puts EI~UMa's magnetic field 
strength at the very high end. Thus, either EI~UMa's magnetic field 
strength is very high for an IP, or the system is an exception to the 
$\zeta \simeq \zeta_{max}$ trend. Interestingly, Figure 7 of \citet
{warner96} shows two other IPs with similarly high magnetic moments and 
mass accretion rates: V1223 Sgr and V1062 Tau, two of the three other 
systems also thought to exhibit mass transfer bursts from the secondary, 
as mentioned in \S5.2. Using the illustrative upper limit for the white 
dwarf mass yields a magnetic field strength of 170~MG, well above the 
typical range for IPs. A magnetic field of this strength is problematic, 
since spectra taken by \citet{thorstensen86} do not show any hint of 
cyclotron harmonics or Zeeman splitting expected with such high $B_1$ 
values. Since higher $B_1$ values correlate with smaller radii and higher 
$M_1$ values, the white dwarf mass is probably be closer to the lower mass 
limit. 

A strong magnetic field moves the inner radius of the disk, given by 
the magnetosphere radius $r_{mag}$, outwards. Given the above parameters 
and eq.~6.5 of \citet{warner95}, we find $r_{mag} > $($1.2 - 1.3$) $\times 
10^{10}$~cm.  There is still room for a disk to form however, 
since the circularization radius $r_{circ} > $($1.7 - 1.8$) $\times 
10^{10}$~cm \citep[eq.~4.20]{frank02} is greater than $r_{mag}$. The 
fastness parameter $P_{Kep} / P_{spin} \gtrsim 1.1$, where $P_{Kep}$ is 
the Keplerian period at $r_{mag}$.  Because $P_{Kep} / P_{spin} > 1$, 
there is likely some propeller component to EI~UMa's accretion. Since the 
$\zeta \approx \zeta_{max}$ assumption brings EI~UMa close to the disk/ 
no disk boundary of Figure 8, accretion in the EI~UMa system may look 
less like a disk and more like an ``accretion ring.''  Given the long 
orbital period, high magnetic moment, and ring-shaped accretion disk, 
we speculate that EI~UMa will eventually evolve into a diskless system 
as the orbital separation slowly decreases. Hence, EI~UMa appears to be 
the precursor to a polar.

\subsection{Hidden Intermediate Polars} %-----------------------

Prior to this work, several researchers suggested that EI~UMa is an
IP (e.g. \citealt{baskill05}, \citealt{pandel05}, 
\citealt{thorstensen86}).  Our discovery of optical pulsations at the spin 
and spin-orbit sideband have confirmed these suspicions. This has led us 
to wonder if the heavy reliance on the detection of a stable 
pulsation for defining an IP may be too conservative, and that we are 
failing to recognize members of the IP class. If true, this would affect 
statistical studies of the IP population as a whole. In the worst case 
scenario, we may be recognizing and studying only the most spectacular 
members of the class, leading to a bias in our understanding of the X-ray 
and optical spin modulation mechanisms.

\citet{baskill06} propose that LS~Peg is such a ``hidden IP''; its X-ray 
spectral characteristics are IP--like, yet it shows no permanent 
pulsation. The authors point out that the degree of misalignment 
between the spin and magnetic axes of the white dwarf is an important
factor in the production of a modulation: if the two axes are nearly 
aligned, any photometric modulations on the spin period would become much 
weaker, perhaps undetectable. V426~Oph is another CV with a highly 
absorbed X-ray spectrum, a characteristic of IPs, and hence is thought to 
be an IP \citep{baskill05}. We note that TW~Pic \citep{norton00}
may be considered a third candidate hidden IP. EI~UMa sometimes 
exhibits a pulse amplitude too low to detect, and could also be called 
a hidden IP.  The existence of four such systems suggests a sizable 
population of hidden IP's may exist, perhaps as many as  15\% (=4/26) of 
all IPs.  The above simple argument hints that the spin and magnetic field 
axes may be aligned in roughly 15\% of cataclysmic variable white dwarfs.

% -----------------------------------------------------------------
\section{Conclusions}

1.) We have detected stable, $\sim$1\% amplitude oscillations in optical 
photometry near the X-ray oscillation period discovered by 
\citet{baskill05}. We attribute a U-band periodicity near 745~s to the 
white dwarf spin, and a V- and R-band periodicity near 770~s to the 
spin--orbit sideband. A periodicity at $\sim$ 813~s is sometimes present 
in the V- and R-bands and may be related to the 3rd spin-orbit sideband 
(expected at $821 \pm 3$~s), but the association is tentative. The 
relative signal amplitude is variable by at least a factor of 2. 

We have also detected a periodicity in archival XMM-Newton observations 
consistent with the 745~s periodicity. The XMM-Newton observations suggest 
the X-ray oscillation is stronger at softer energies ($< 2$~keV) than at 
hard energies. The U-band periodicity agrees with both X-ray periodicity 
detections, and combined give a spin period of $745.7 \pm 2.7$~s. Combined 
with other evidence from previous X-ray observations (\citealt{pandel05}; 
\citealt{baskill05}), our observations strongly suggest that EI~UMa is an 
intermediate polar. 

2.) An outburst event was observed in 2005 November -- December with an 
amplitude $> 0.65$ magnitudes in the R-band and a full duration of at 
least 20 days. The decline of a flare on top of the longer timescale 
outburst event was observed on 2005 Nov 20, in which the R-band brightness 
dropped by 0.27~mag in 3 hours. The low amplitude and long duration of 
the outburst are uncharacteristic of normal dwarf novae.  We argue that 
it is unlikely that EI~UMa is a dwarf nova, a conclusion supported by 
AAVSO and VSNet observations going back to 1993 and by the high mass 
transfer rate we derive. By analogy with TV Col, we interpret the 
outburst and the flare as mass transfer events from the secondary star.

3.) Using the Barnes-Evans surface brightness relation, we estimate the 
distance to EI~UMa is $750^{+100}_{-200}$ pc. The disk mass transfer 
rate $\dot{M} \simeq 3.6 \times 10^{17}$ g s$^{-1}$, likely greater than 
the critical rate above which dwarf nova outbursts are prevented. 
Assuming EI~UMa exhibits typical IP behavior, the magnetic moment is 
$(3.4 \pm 0.2) \times 10^{33}$~G cm$^3$. This strong magnetic moment, 
paired with a long orbital period, implies that EI~UMa contains an 
accretion ring rather than a disk.  We speculate that EI~UMa is a 
pre-polar and will evolve into a diskless cataclysmic variable.

% -----------------------------------------------------------------
\section{Acknowledgments}

We thank Nassissie Fedaku and Danny Martino for assisting with the 
observations in 2003, and Allen Shafter and the referee for valuable 
comments.  We also gratefully acknowledge the variable star observers 
with the AAVSO and the VSNet, whose long-term observations were used 
in this research.  This 
research has made use of NASA's Astrophysics Data System and the SIMBAD 
database, operated at CDS, Strasbourg, France. IRAF is distributed by the 
National Optical Astronomy Observatories, which are operated by the 
Association of Universities for Research in Astronomy, Inc., under 
cooperative agreement with the National Science Foundation. Support for 
this research was provided by NASA through grants HST-GO-09078.01A and 
HST-GO-08156.02A from the Space Telescope Science Institute, which is 
operated by the Association of Universities for Research in Astronomy, 
Incorporated, under NASA contract NAS5-26555.

% -----------------------------------------------------------------

%----------------------------------------------------------------
\clearpage

\input{tab1.tex}

%----------------------------------------------------------------
\clearpage

\input{tab2.tex}

%----------------------------------------------------------------
\clearpage

\input{tab3.tex}

%----------------------------------------------------------------
\clearpage
\begin{figure}
\epsscale{.75}
\plotone{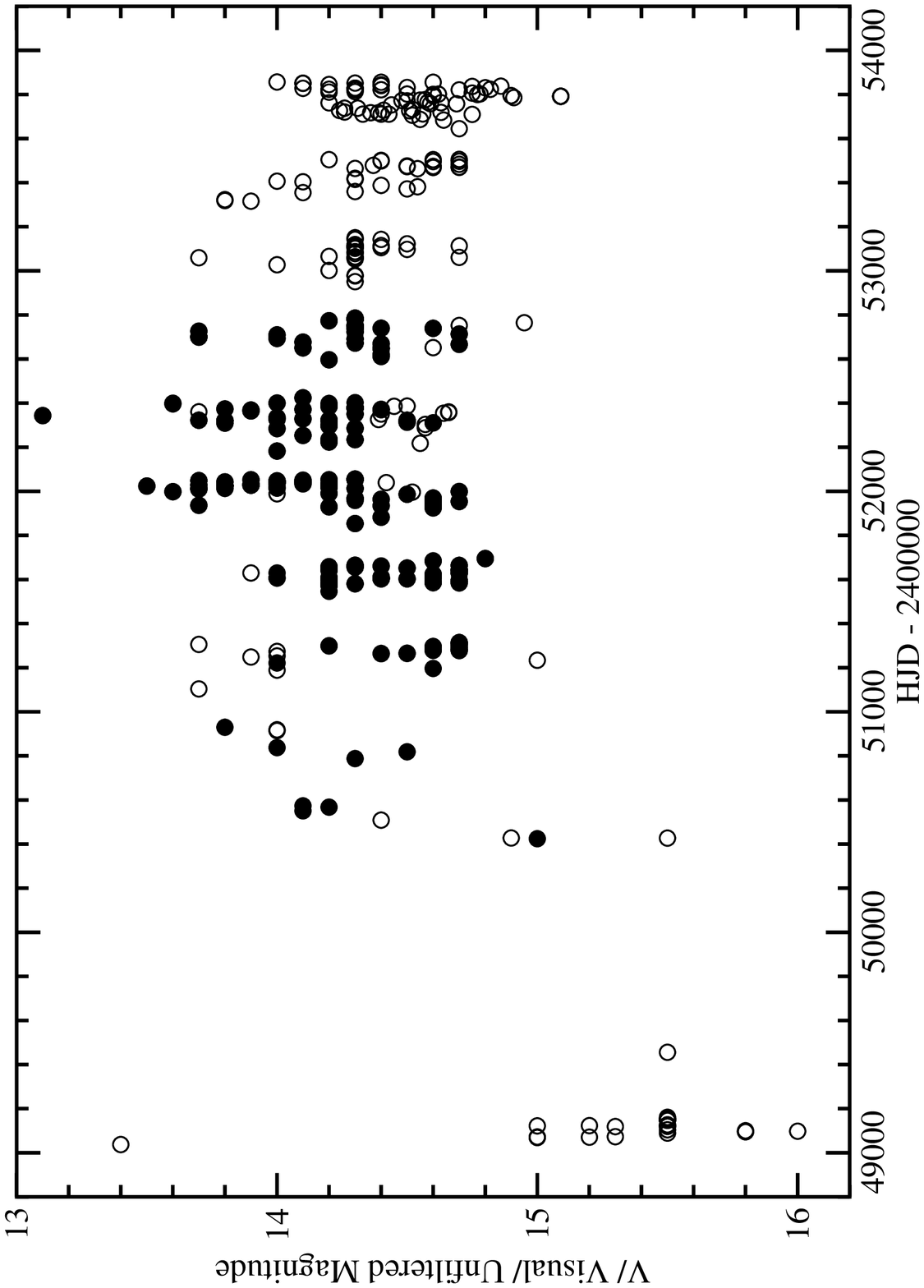}
\caption{
Light curve of EI UMa from 1993 to 2006, as provided by the AAVSO 
(open circles) and the VSNet (solid circles; \citealt{katonogami03}). 
Note the lack of dwarf nova outbursts. 
\label{fig1}}
\end{figure}

%----------------------------------------------------------------
\clearpage

\begin{figure}
\epsscale{.75}
\plotone{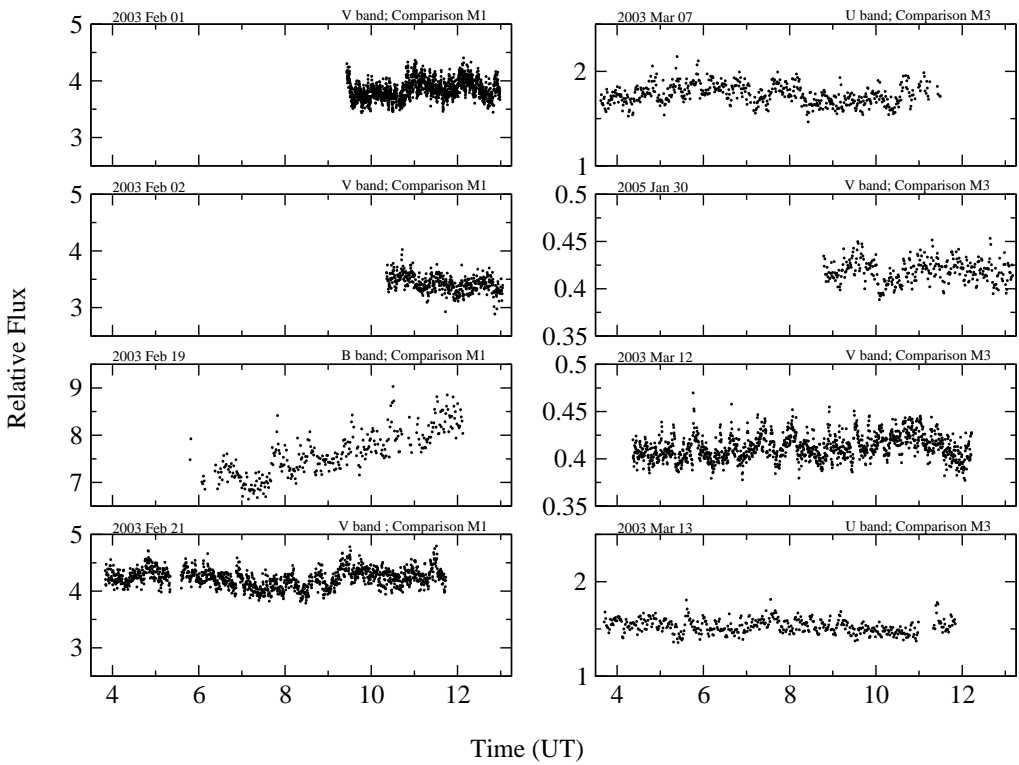}
\caption{
U, B, and V band light curves of EI UMa. The time is in hours, and the 
flux is relative to comparison stars from \citet{misselt96}. The most 
notable characteristic is the random flickering, a few percent in 
amplitude.
\label{fig2}}
\end{figure}

%----------------------------------------------------------------
\clearpage

\begin{figure}
%\epsscale{.5}
\plotone{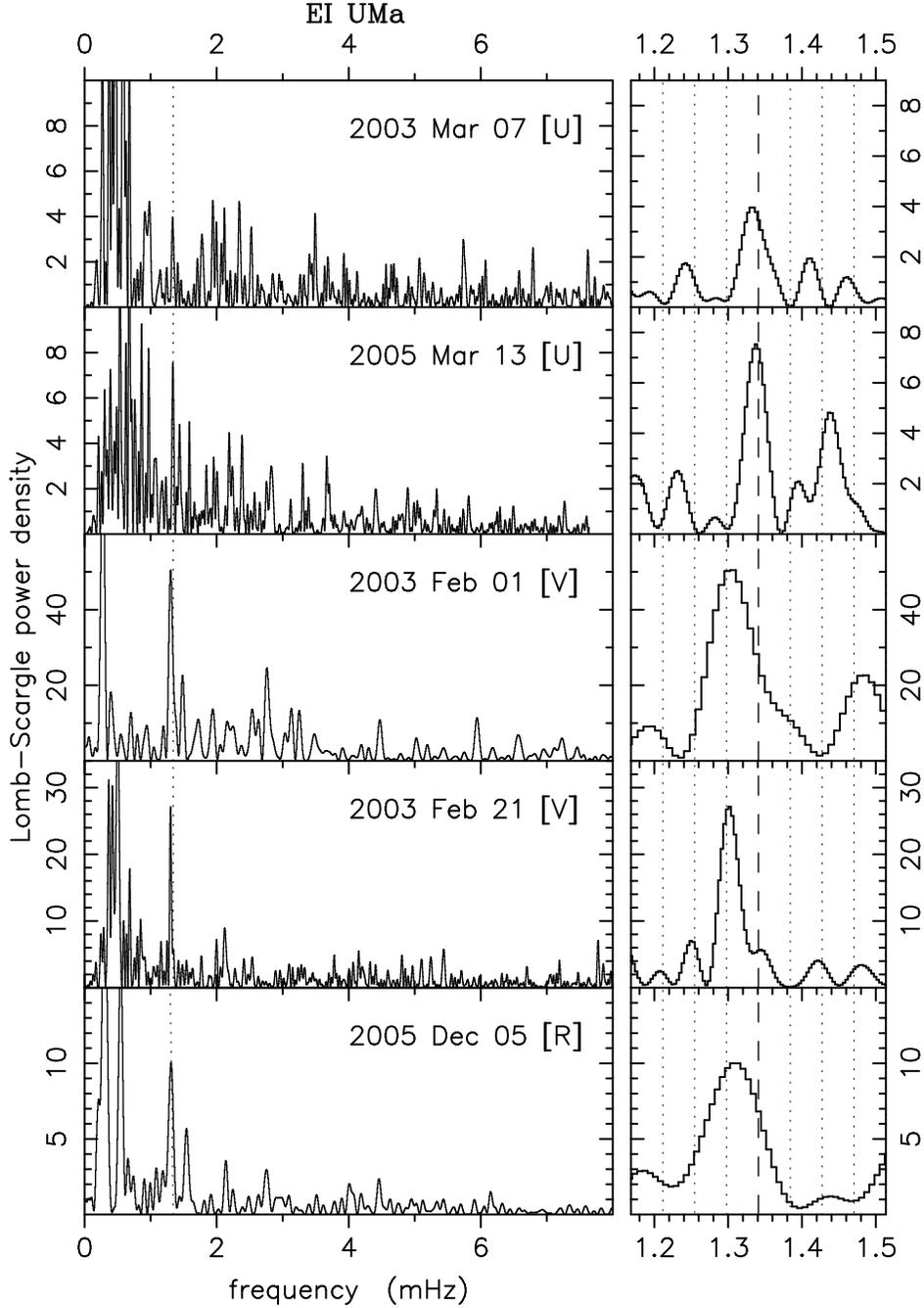}
\caption{
Power spectra of prewhitened optical light curves, 10x oversampled, 
showing spin and first sideband periods signals. U-band power spectra 
in the upper two panels show signals at the 745~s spin period, while 
the V- and R-band power spectra in the lower panels show signals 
consistent with the first sideband, at 770~s. The left hand panels 
show the low-frequency part of the power spectra and the right-hand 
panels zoom in on the region between 680 and 825~s. 
A dashed line marks the spin period and dotted lines mark the upper and 
lower spin-orbit sidebands. 
\label{fig3}}
\end{figure}

%----------------------------------------------------------------
\clearpage

\begin{figure}
%\epsscale{.5}
\plotone{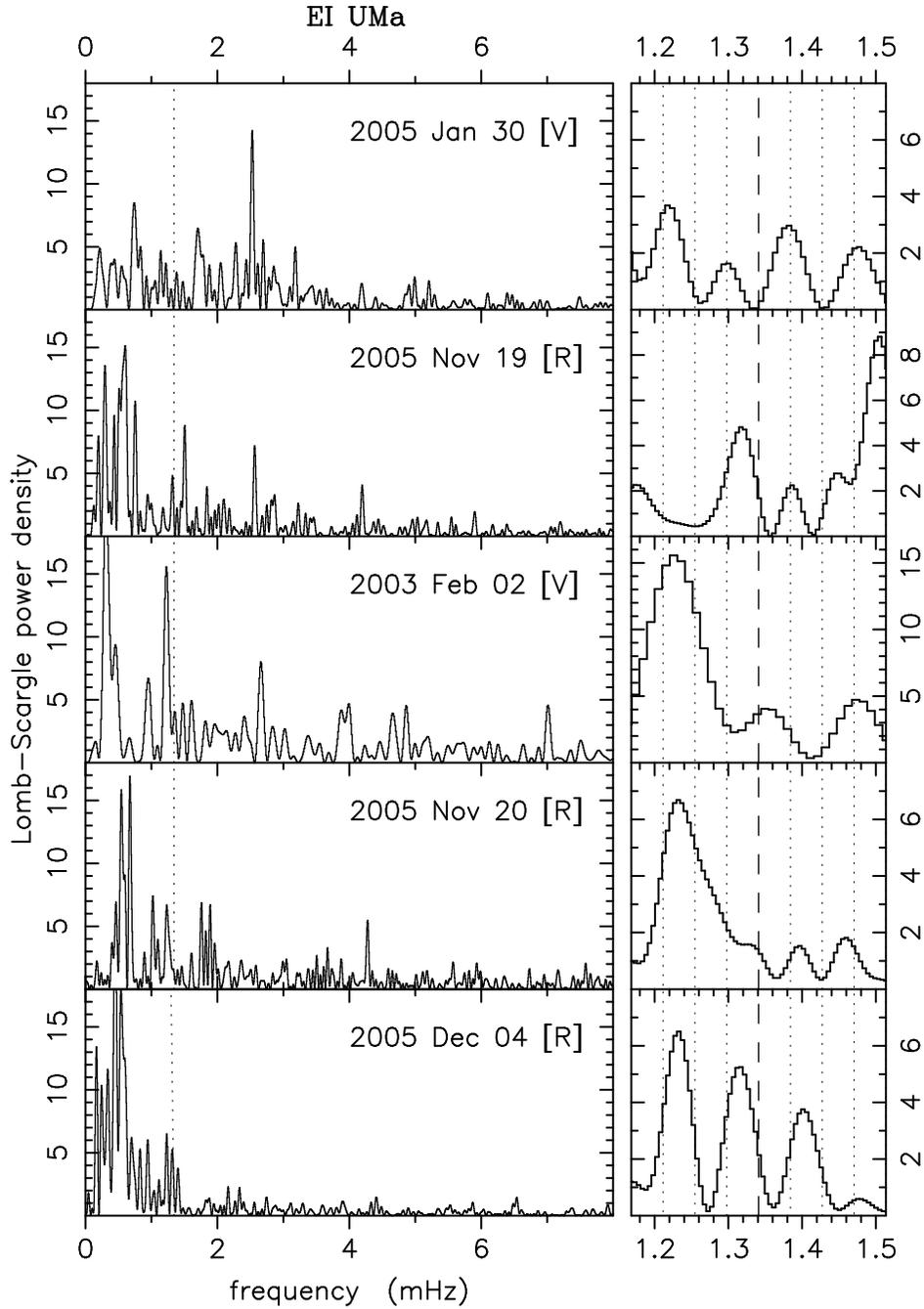}
\caption{
Power spectra of prewhitened V and R band light curves. The two upper 
panels show examples of non-detections, while the three bottom panels show 
possible detections of a third-order sideband, particularly on 2003 Feb 02.
Compare with Fig.~3. 
\label{fig4}}
\end{figure}

%----------------------------------------------------------------
\clearpage

\begin{figure}
%\epsscale{.5}
\plotone{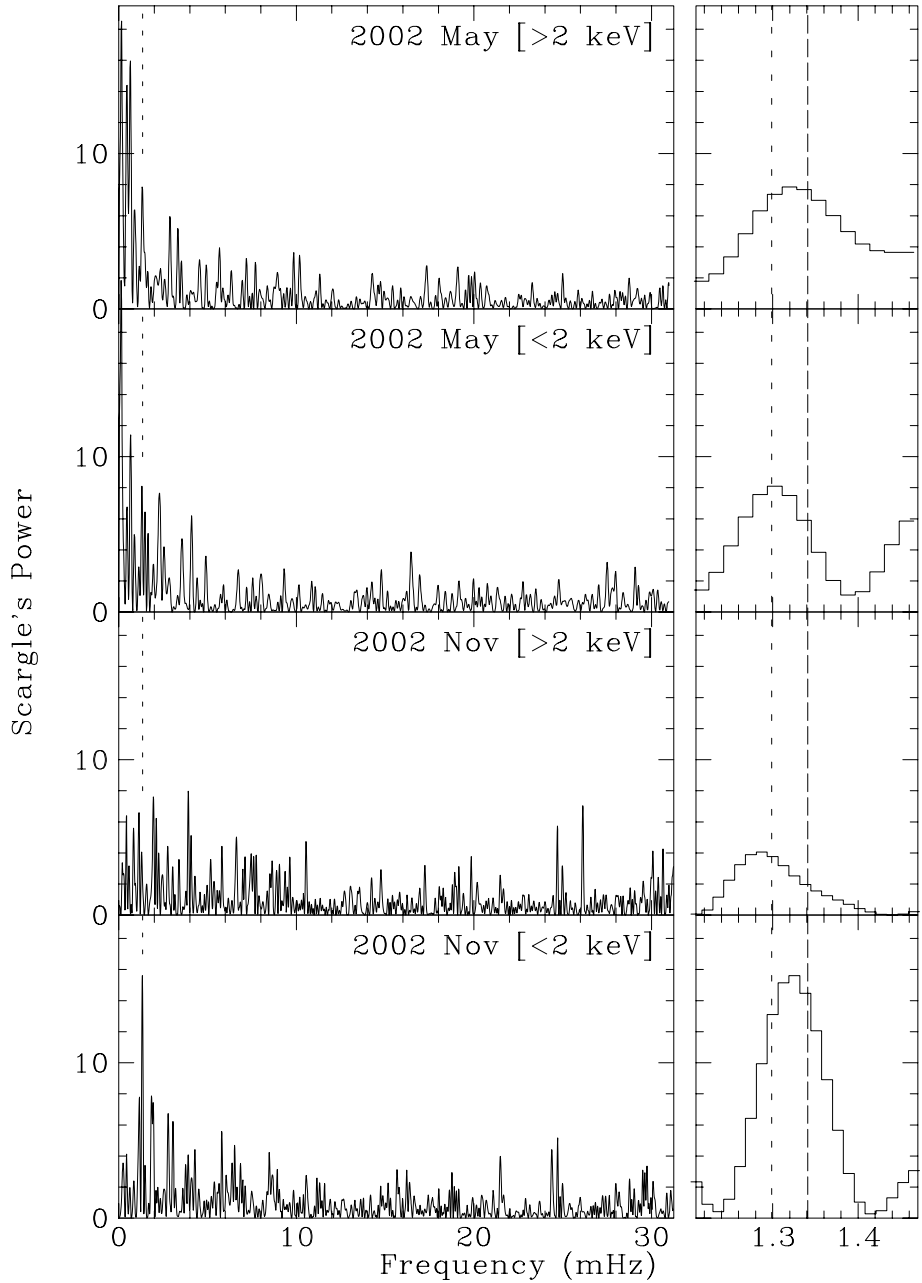}
\caption{
Power spectra of XMM-Newton X-ray light curves taken at two epochs and 
separated into hard and soft bandpasses. The right-hand panels 
show the region between 680 and 825~s (equal to the $\pm 3$ spin-orbit 
sideband). The dashed line marks the spin period and the dotted line 
marks the first spin-orbit sideband.
\label{fig5}}
\end{figure}

%----------------------------------------------------------------
\clearpage

\begin{figure}
%\epsscale{.5}
\plotone{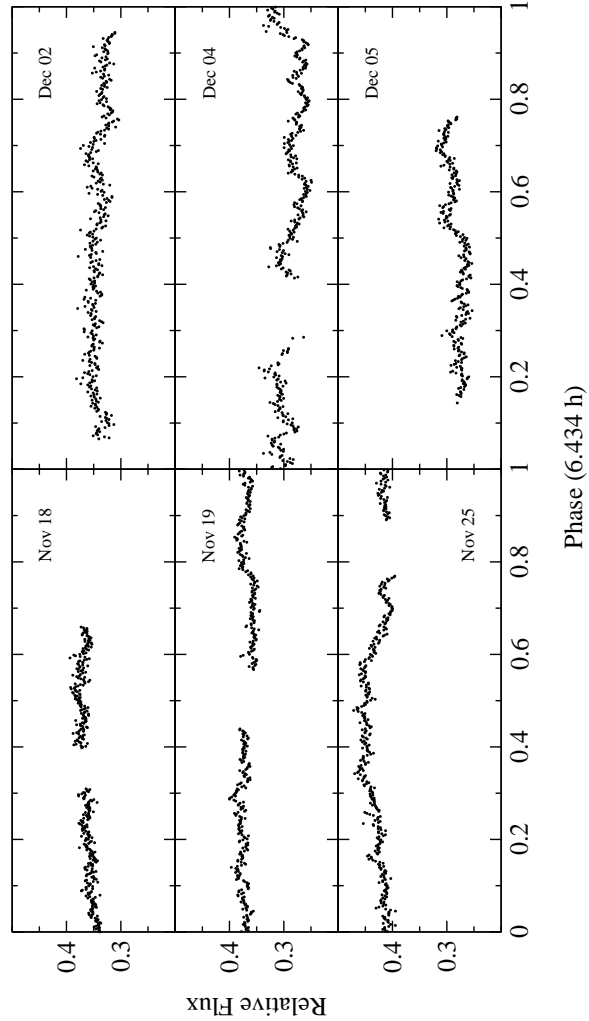}
\caption{
R-band light curves folded on the 6.434~h orbital period.  The flux is 
relative to comparison star M3 \citep{misselt96}. No obvious periodic 
modulation on the orbital period is present. In particular, there is
no evidence of any bright-spot emission or ellipsoidal variation, 
suggesting the system has a low orbital inclination, a high-mass transfer 
rate, or both.
\label{fig6}}
\end{figure}

%----------------------------------------------------------------
\clearpage

\begin{figure}
%\epsscale{.5}
\plotone{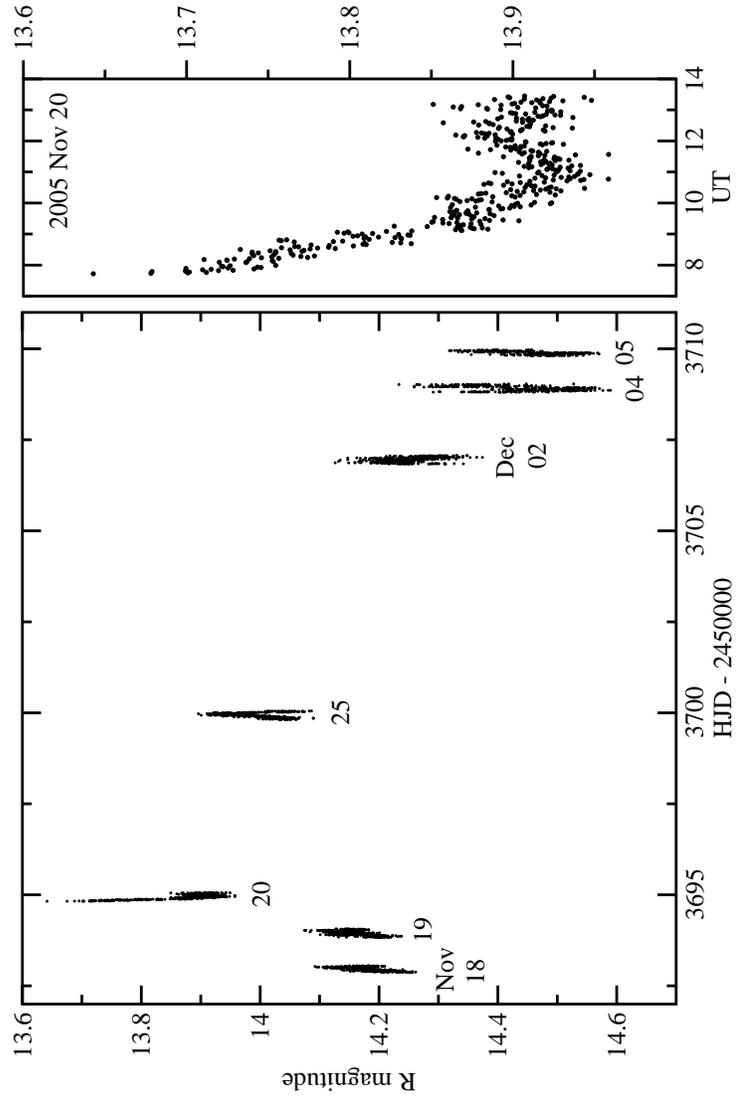}
\caption{
R-band light curve of the 2005 November/ December outburst.  Though 
unresolved, the outburst lasted $\sim$20~d. Inset: a magnified view of 
the 2005 November 20 light curve showing the rapid decay of a flare.
\label{fig7}}
\end{figure}

%----------------------------------------------------------------
\clearpage

\begin{figure}
\plotone{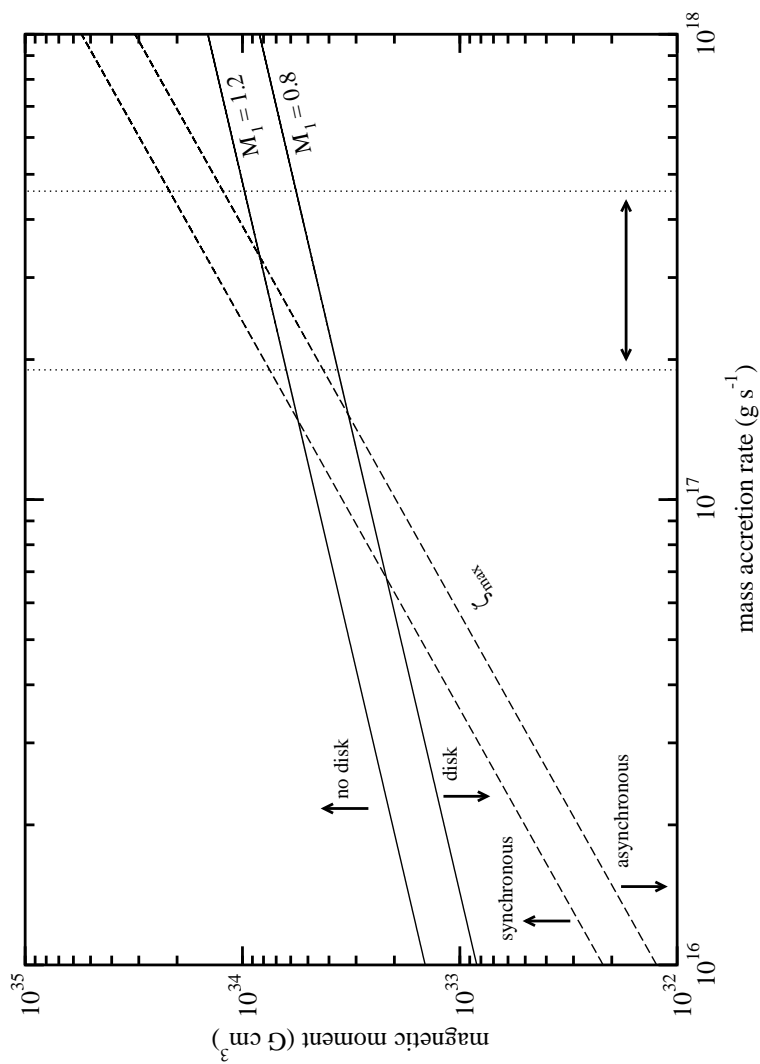}
\caption{
Magnetic moment versus mass accretion rate plot for $P_{orb} = 6.434$~h, 
and for $M = 0.8 \Msun$ or $1.2 \Msun$.  The disk condition is marked by 
solid lines, while the synchronization condition is marked by dashed 
lines. The $\dot{M}$ range for EI UMa is bounded by the vertical dotted 
lines.
\label{fig8}}
\end{figure}

%----------------------------------------------------------------

\end{document}

%% file: tab1.tex
Table 1: Observation log\\
\begin{tabular}{c|cccccc}
        \hline
Night&Filter&UT Start&Duration& Aperture & Seeing   & Comparison\\
     &      &Time    &(h:mm)  & (arcsec) & (arcsec) & Star\\
        \hline
2003 Feb 01 & V & 09:26 & 3:33 & 1.6 & 1.0 & M1\\
2003 Feb 02 & V & 10:21 & 2:41 & 2.0 & 3.1 & M1\\
2003 Feb 19 & B & 05:48 & 6:19 & 1.6 & 1.2 & M1\\
2003 Feb 21 & V & 03:50 & 7:53 & 2.0 & 2.6 & M1\\
2003 Mar 07 & U & 03:38 & 7:52 & 2.0 & 1.3 & M3\\
2005 Jan 30 & V & 08:47 & 4:23 & 1.6 & 1.2 & M3\\
2005 Mar 12 & V & 04:22 & 7:51 & 1.6 & 1.4 & M3\\
2005 Mar 13 & U & 03:42 & 8:09 & 2.4 & 1.9 & M3\\
2005 Jun 07 & B & 04:50 & 1:18 & 2.0 & 2.4 & M3\\
	\hline
2005 Nov 18 & R & 08:55 & 4:14 & 4.8 & 2.2 & M3\\
2005 Nov 19 & R & 07:52 & 5:39 & 4.8 & 3.1 & M3\\
2005 Nov 20 & R & 07:44 & 5:43 & 4.8 & 2.6 & M3\\
2005 Nov 25 & R & 07:29 & 6:25 & 4.8 & 3.0 & M3\\
2005 Dec 02 & R & 07:54 & 5:39 & 4.8 & 2.7 & M3\\
2005 Dec 04 & R & 07:10 & 6:22 & 4.8 & 4.8 & M3\\
2005 Dec 05 & R & 07:10 & 3:58 & 4.8 & 3.5 & M3\\
        \hline
\end{tabular}\\

%% file: tab2.tex
Table 2: Observed properties of EI UMa\\
\begin{tabular}{c|ccccc}
        \hline
Night & Filter & Mean      & RMS             & Period   & Amplitude\\
      &        & Magnitude & Flickering (\%) & (seconds)& (\%)\\
        \hline
2003 Feb 01 & V & 14.58 & 4.4 & $770\pm3$     & $1.3\pm0.2$ \\
2003 Feb 02 & V & 14.70 & 4.3 & $813\pm7$     & $1.2\pm0.2$ \\
2003 Feb 19 & B & 14.64 & 6.4 & -             & $<0.7$ \\
2003 Feb 21 & V & 14.48 & 3.8 & $768\pm2$     & $0.9\pm0.1$ \\
2003 Mar 07 & U & -     & 6.0 & $748\pm5$     & $0.7\pm0.3$ \\
2005 Jan 30 & V & 14.47 & 2.9 & -             & $<0.6$ \\
2005 Mar 12 & V & 14.48 & 3.2 & $765\pm3$     & $0.5\pm0.1$ \\
2005 Mar 13 & U & -     & 4.8 & $747\pm4$     & $1.0\pm0.3$ \\
2005 Jun 07 & B & 15.00 & 3.5 & -             & $<1.0$ \\
        \hline 
2005 Nov 18 & R & 14.18 & 3.1 & -             & $<0.6$ \\
2005 Nov 19 & R & 14.16 & 2.9 & -             & $<0.5$ \\
2005 Nov 20 & R & 13.87 & 6.1 & $812\pm7$     & $0.4\pm0.1$ \\
2005 Nov 25 & R & 13.99 & 4.1 & -             & $<0.5$ \\
2005 Dec 02 & R & 14.24 & 4.1 & -             & $<0.8$ \\
2005 Dec 04 & R & 14.44 & 7.2 & $815\pm5$     & $1.0\pm0.4$ \\
2005 Dec 05 & R & 14.46 & 5.4 & $760\pm6$     & $1.1\pm0.3$ \\
        \hline
\end{tabular}

%% file: tab3.tex
Table 3: Outburst Properties for Intermediate Polar Dwarf Nova 
Candidates\\
\begin{tabular}{c|cccc}
\hline
System      &$P_{orb}$ & Duration   & $\Delta m$    & Recurrance\\
            &(h)       & (d)        &               & Interval\\
\hline
HT Cam      & 1.43     & $\sim 2$   & $\sim 4.5$    & $\sim$ 150 d?\\
EX Hya      & 1.64     & $2 - 3$    & $\sim 3.5$    & $\sim$ years\\
V1223 Sgr   & 3.37     & $\sim 0.5$ & $>1$          & ?\\
YY (DO) Dra & 3.96     & $\sim 5$   & $\sim 5$      & $\sim 1000$ d\\
CW Mon      & 4.23     & $\sim 20$  & $\sim 2$      & 100 - 200 d\\
TV Col      & 5.49     & $\sim 0.5$ & $\sim 2$      & $\sim$ 1 month?\\
XY Ari      & 6.06     & $\sim 5$   & $\sim 3$      & $\gtrsim 50$ d\\
EI UMa      & 6.43     & $\lesssim 1$  & $>0.8$     & ?\\
V1062 Tau   & 9.95     & $\sim 1-2$ & $\sim 1.2$  & $\lesssim 6$ months\\
GK Per      & 47.9     & $\sim 30 - 70$ & $\sim 3$  & 880-1240 d\\
\hline
\end{tabular}
\\
References. General: \citet{hellier97}. TV Col:  
\citet{hellierbuckley93}. V1223 Sgr: \citet{amerongen89}. V1062 Tau: 
\citet{lipkin04}. HT Cam: \citet{ishioka02}. EX Hya: 
\citet{hellier00}. YY Dra: \citet{wenzel83}. XY Ari: 
\citet{hellier97}. CW Mon: \citet{kato03}. GK Per: 
\citet{simon02}.